\documentclass{article}
\usepackage{xcolor} 
\usepackage{amsmath}
\usepackage[top=0.6in, left=1.7in, right=1.7in]{geometry}
\usepackage{graphicx}
\usepackage{parskip}
\usepackage{subcaption}
\usepackage{float} 
\usepackage{cleveref}
\usepackage{soul}
\usepackage{array}
\usepackage{natbib}
\usepackage{chngcntr}      
\counterwithin{figure}{section}
\title{Auditing LLMs for Algorithmic Fairness in Casenote-Augmented Tabular Prediction\vspace{-1em}}
\usepackage{fullpage}
\author{}
\date{}
\begin{document}
\maketitle

\begin{center}
    \begin{minipage}{0.4\textwidth}
        \centering
        \textbf{Xiao Qi Lee} \\
        Department of Computer Science \\
        San Jose State University \\
        \texttt{xiaoqi.lee@sjsu.edu}
    \end{minipage}
    \hfill
    \begin{minipage}{0.4\textwidth}
        \centering
        \textbf{Ezinne Nwankwo} \\
        Department of Computer Science \\
        University of California, Berkeley \\
        \texttt{ezinne\_nwankwo@berkeley.edu}
    \end{minipage}

    \vspace{1em} 

    \begin{minipage}{0.6\textwidth}
        \centering
        \textbf{Angela Zhou} \\
        Department of Data Science and Operations \\
        University of Southern California \\
        \texttt{zhoua@usc.edu}
    \end{minipage}
\end{center}
\vspace{1em}

\begin{abstract}
    LLMs are increasingly being considered for prediction tasks in high-stakes social service settings, but their algorithmic fairness properties in this context are poorly understood.
In this short technical report, we audit the algorithmic fairness of LLM-based tabular classification on a real housing placement prediction task, augmented with street outreach casenotes from a nonprofit partner. 
We audit multi-class classification error disparities. 
We find that a fine-tuned model augmented with casenote summaries can improve accuracy while reducing algorithmic fairness disparities. We experiment with variable importance improvements to zero-shot tabular classification and find mixed results on resulting algorithmic fairness.
Overall, given historical inequities in housing placement, it is crucial to audit LLM use. We find that leveraging LLMs to augment tabular classification with casenote summaries can safely leverage additional text information at low implementation burden. The outreach casenotes are fairly short and heavily redacted. Our assessment is that LLM zero-shot classification does not introduce additional textual biases beyond algorithmic biases in tabular classification. Combining fine-tuning and leveraging casenote summaries can improve accuracy and algorithmic fairness.
\end{abstract}

\section{Introduction}

Large language models, and other natural language processing tools trained on word embeddings and other text corpora, have been widely demonstrated to be biased along gender, race, religion, and other protected attributes. Such bias can be taxonomized in different ways; from common analogical tests used to highlight the promulgation of stereotypes in word embeddings, to audit-type tests that illustrate different meritorious judgments that differ on race/gender signals alone, to representational biases that arise when generating language anew. How and whether to mitigate such societal biases remains an open question. At the same time, tabular (non-LLM-based) prediction has also been widely demonstrated to exhibit potentially disparate error behaviors for different protected attributes for a variety of reasons: different group sizes, different underlying heterogeneity, different base rates or noise rates. It stands to reason that leveraging LLMs for prediction may be subject to similar algorithmic fairness concerns as tabular prediction is, if not more. The deployment of LLMs in consequential domains must therefore proceed with caution and should incorporate algorithmic fairness evaluations. 

We are motivated by an ongoing collaboration with a nonprofit that conducts street outreach for homelessness services. During street outreach, outreach workers record casenotes after every interaction with a client --- therefore, these casenote records comprise the richest information about individual clients. However, outreach interactions are unstructured and open-ended. There are so many casenotes that it is generally impossible to view these as data or a source of analytical information, beyond inspecting individual's records. Large language models can be particularly helpful in parsing unstructured data, such as these casenote histories, for the purposes of prediction or extracting structured information. 

Nonetheless, before investigating the potential use cases for LLMs for social service casenotes, it is important to investigate whether leveraging LLMs for prediction could introduce algorithmic bias. There is a long history of racial discrimination in housing decisions; discriminatory actions by landlords can affect housing placements. Although we work with redacted casenotes, which generally remove fine-grained locational information or any names, other sources of algorithmic bias could persist, such as differences in minority/majority group size or different housing patterns. In this work, we conduct an empirical audit of LLM-based tabular classification in this setting, examining both predictive performance and algorithmic fairness across demographic subgroups.

The contributions of this brief technical report are the following. \textbf{(1)} We audit zero-shot and fine-tuned LLM performance on a tabular housing-placement prediction task derived from real social-service casenotes, benchmarking 17B and 70B Llama variants against a random forest baseline. \textbf{(2)} We evaluate algorithmic fairness using multiclass Statistical Parity and Equality of Opportunity across race and gender subgroups, and observe that imbalanced fine-tuning data can amplify subgroup disparities even as overall accuracy improves. \textbf{(3)} We investigate the effects of two prompt-augmentation strategies---casenote summaries and feature-importance cues---showing that summaries can reduce fairness gaps even when they do not improve accuracy, while emphasizing a single historically predictive feature can reinforce historical bias. Together, these findings offer practical guidance for deploying LLMs responsibly in high-stakes social-service settings.

\section{Methodology}
\subsection{Prediction task: Serialized tabular classification.}

We study housing placement prediction for clients served by a nonprofit street outreach organization, using the same dataset introduced in \citet{nwankwo2025batch}. The dataset consists of 471 clients seen consistently at least once per month from 2019--2021. The outcome $Y \in \{0,1,2,3\}$ is the highest housing placement level each client achieved by 2021, where 0 indicates remaining on the streets, 1 indicates other temporary placements (e.g., hospital), 2 indicates shelter or transitional housing, and 3 indicates permanent supportive housing.

Each client is described by a set of tabular covariates recorded at baseline, including: date first seen by the outreach team, outreach program, perceived chronic homelessness status, gender, race, ethnicity, age range, whether the contact was initiated by a 311 or 911 call, whether a removal hotline call was involved, and casenote-extracted binary indicators for housing application activity, service refusal, possession of important documents, and receipt of social service benefits. Additional covariates include the number of outreach engagements prior to 2019 and the maximum housing placement reached before 2019---the latter was identified as the most predictive feature via random forest feature importances.

For LLM-based classification, the tabular record for each client is serialized into a natural-language prompt describing their covariates and, in augmented variants, an LLM-generated summary of their casenotes recorded prior to the prediction window. The model is then asked zero-shot or after fine-tuning to predict the client's most likely housing placement level (0--3). We evaluate Llama 3 models at 17B and 70B parameter scales, using the Together AI API for enterprise-secure inference, and compare against a random forest trained on the tabular covariates alone.

\subsection{Fairness Evaluation Metrics}
Because housing placement is a multiclass outcome, standard binary fairness metrics do not directly apply. We use multiclass extensions of two standard metrics: Statistical Parity (\colorbox{pink}{SP}) and Equality of Opportunity (\colorbox{yellow}{EoO}). To avoid unstable estimates from small cells, we restrict evaluation to subgroups with sufficient sample sizes, specifically race (White, Black, Other) and gender (Male, Female).

Here $Y$ denotes the true class label, $\hat{Y}$ the predicted label, $c \in \{0,\dots,C-1\}$ a class, $A$ a protected attribute, and $g$ a subgroup within $A$.

\textbf{Statistical Parity} measures whether predicted outcomes are independent of group membership. For each group $g$ and class $c$, we compute the absolute deviation from the marginal prediction rate:
$$d_{\colorbox{pink}{\text{SP}}}(c, g) = \left| P(\hat{Y} = c \mid A = g) - P(\hat{Y} = c) \right|$$
and summarize across classes as:
$$\text{SP Mean}(g) = \frac{1}{C} \sum_{c=0}^{C-1} d_{\colorbox{pink}{\text{SP}}}(c, g), \qquad \text{SP Max}(g) = \max_c\, d_{\colorbox{pink}{\text{SP}}}(c, g).$$

\textbf{Equality of Opportunity} measures whether groups achieve equal true positive rates (TPR) for each class. Let $\text{TPR}_c(g) = P(\hat{Y} = c \mid Y = c, A = g)$ and $\text{TPR}_c = P(\hat{Y} = c \mid Y = c)$. Then:
$$d_{\colorbox{yellow}{\text{EoO}}}(c, g) = \left| \text{TPR}_c(g) - \text{TPR}_c \right|$$
$$\text{EoO Mean}(g) = \frac{1}{C'} \sum_{c=0}^{C'-1} d_{\colorbox{yellow}{\text{EoO}}}(c, g), \qquad \text{EoO Max}(g) = \max_c\, d_{\colorbox{yellow}{\text{EoO}}}(c, g)$$
where $C'$ is the number of classes with at least one true occurrence; classes with no true occurrences yield undefined TPR and are excluded.

\vspace{-0.5em}
\section{Results}
\subsection{Models Without Summaries}

Tables 1–3 present the performance of all model variants, tested in zero-shot and fine-tuned settings without summaries, using a 75–10–15 train–validation–test split. The corresponding confusion matrices are provided in the Appendix.

\begin{table}[H]
    \vspace{-0.5em} 
    \centering
    \normalsize
    \renewcommand{\arraystretch}{1.1}
    \setlength{\tabcolsep}{2pt}
    \caption{Evaluation Metrics for Zero-Shot 17B Model}
    \begin{tabular}{lcccccccc}
        \hline
        Model & Accuracy & F1 Score & RMSE & RMSE \% & SP Mean & SP Max & EOO Mean & EOO Max \\
        \hline
        zs\_all    & 0.331 & 0.191 & 1.282 & 57.27 & 0.025 & 0.094 & 0.028 & 0.088 \\
        \hline
    \end{tabular}
\end{table}

\begin{table}[H]
    \vspace{-0.5em} 
    \centering
    \normalsize
    \renewcommand{\arraystretch}{1.1}
    \setlength{\tabcolsep}{2pt}
    \caption{Evaluation Metrics for Zero-Shot 70B Model}
    \begin{tabular}{lcccccccc}
        \hline
        Model & Accuracy & F1 Score & RMSE & RMSE \% & SP Mean & SP Max & EoO Mean & EoO Max \\
        \hline
        zs\_all & 0.232 & 0.164 & 1.140 & 62.01 & 0.023 & 0.079 & 0.012 & 0.112 \\
        \hline
    \end{tabular}
\end{table}

\begin{table}[H]
    \vspace{-0.5em} 
    \centering
    \normalsize
    \renewcommand{\arraystretch}{1.1}
    \setlength{\tabcolsep}{2pt}
    \caption{Evaluation Metrics for Fine-Tuned 70B Model}
    \label{tab:ft_metrics}
    \begin{tabular}{lcccccccc}
        \hline
        Model & Accuracy & F1 Score & RMSE & RMSE \% & SP Mean & SP Max & EoO Mean & EoO Max \\
        \hline
        ft\_all & 0.702 & 0.546 & 1.254 & 58.19 & 0.027 & 0.110 & 0.071 & 0.209 \\
        \hline
    \end{tabular}
\end{table}

\vspace{-0.8em} 

\textbf{Conclusions: } Without summaries, the fine-tuned 70B achieves the strongest accuracy (70.2\%), while both 17B and 70B zero-shot models perform signficantly worse. However, this increase in  performance comes with increased fairness disparities as shown through higher EoO mean and max values. This highlights a clear performance–fairness tradeoff: fine-tuning improves accuracy but amplifies subgroup disparities.

\subsection{Models With Summaries}
Using the base prompt, casenote summary information was added to the models and then tested on 17B and 70B models in a zero-shot setting. Tables 4-6 show their results accordingly.

\begin{table}[H]
    \vspace{-0.5em} 
    \normalsize
    \renewcommand{\arraystretch}{1.1}
    \setlength{\tabcolsep}{2pt}
    \centering
    \caption{Evaluation Metrics for Zero-Shot 17B + Summaries }
    \begin{tabular}{lcccccccc}
        \hline
        Model & Accuracy & F1 Score & RMSE & RMSE \% & SP Mean & SP Max & EoO Mean & EoO Max \\
        \hline
        zs\_all & 0.470 & 0.331 & 1.160 & 61.32 & 0.026 & 0.074 & 0.044 & 0.125 \\
        \hline
    \end{tabular}
\end{table}

\begin{table}[H]
    \vspace{-0.5em} 
    \centering
    \normalsize
    \renewcommand{\arraystretch}{1.1}
    \setlength{\tabcolsep}{2pt}
    \caption{Evaluation Metrics for Zero-Shot 70B + Summaries}
    \begin{tabular}{lcccccccc}
        \hline
        Model & Accuracy & F1 Score & RMSE & RMSE \% & SP Mean & SP Max & EoO Mean & EoO Max \\
        \hline
        zs\_all & 0.308 & 0.259 & 1.122 & 62.59 & 0.030 & 0.092 & 0.039 & 0.109 \\
        \hline
    \end{tabular}
\end{table}

\begin{table}[H]
    \vspace{-0.5em} 
    \centering
    \normalsize
    \renewcommand{\arraystretch}{1.1}
    \setlength{\tabcolsep}{2pt}
    \caption{Evaluation Metrics for Fine-Tuned 70B + Summaries}
    \begin{tabular}{lcccccccc}
        \hline
        Model & Accuracy & F1 Score & RMSE & RMSE \% & SP Mean & SP Max & EoO Mean & EoO Max \\
        \hline
        ft\_all & 0.686 & 0.504 & 1.252 & 58.26 & 0.025 & 0.081 & 0.021 & 0.097 \\
        \hline
    \end{tabular}
\end{table}

Based on the results summarized in Tables 1–6, there are several observations to note. One of which being that incorporating case note summaries into the zero-shot models resulted in accuracy improvements ranging from 7.6\% to 13.9\%. This improvement is likely because zero-shot models rely exclusively on pre-training, so the summaries can help highlight important context and bridge gaps in understanding. However, the addition of summaries had limited effect on the fine-tuned 70B model. Since the fine-tuned model has already been trained on ground‑truth outcome labels for this task, additional summaries may introduce noise that reduces the model’s effectiveness. 

Nevertheless, while the fine-tuned 70B with summaries achieved slightly lower accuracy than the version without summaries, fairness improved substantially, with both EoO mean and EoO max decreasing. This suggests that summaries may still play a useful role in reducing disparities even when they do not meaningfully boost predictive performance.

Furthermore, the zero-shot 17B with summaries outperformed the 70B with summaries, achieving both higher accuracy and lower EoO mean, though with a slightly higher EoO max. This points towards the conclusion that the smaller model was less sensitive to the noise introduced by unstructured case note inputs, resulting in more consistent fairness on average. However, after fine-tuning, the 70B with summaries clearly outperformed the 17B with summaries, showing that larger models are more capable of effectively leveraging nuances in the contextual information to translate their flexibility into stronger predictive performance and improved fairness.

Finally, the fine-tuned 70B model achieved the highest overall accuracy, outperforming the zero-shot models by as much as 37.1\%. However, it also showed the largest EoO mean and maximum disparities. Given that the dataset contains an approximate 4:1 ratio of men to women, fine-tuning appears to have amplified disparities affecting minority groups.\footnote{Beyond the experiments outlined above, we also explored: few-shot prompting on 17B/70B models with 3–5 examples, alternative train-validation-test splits (70–10–20 and 80–10–10), comparing metrics and reasoning traces with protected attributes included or excluded (both with and without summaries), manual prompt optimization, prompt optimization with llama-prompt-ops, running with DateFirstSeen in calendar form vs ordinal format, and SMOTE resampling on tabular data (oversampling females to $\sim$70\% of male rows before serialization). However, none of these consistently improved accuracy or fairness}

\vspace{0.5em}
\textbf{Conclusions:} In the above tables, we see that the addition of case note summaries improved zero-shot accuracy by up to 13.9\% with limited effects on accuracy for the fine-tuned model. Interestingly, the zero-shot 17B with summaries outperforms the 70B with summaries on both accuracy and average fairness, showing smaller models can sometimes handle noisy contextual inputs better. However, after fine-tuning with summaries, 70B's fairness improves significantly.

\section{Random Forest Baseline Results}

To compare the LLM model variants discussed thus far, we also perform the prediction task using a traditional random forest model. The table below summarizes the results.

\begin{table}[ht]
    \centering
    \captionsetup{skip=3pt}
    \normalsize
    \renewcommand{\arraystretch}{1.1}
    \setlength{\tabcolsep}{2pt}
    \caption{Random Forest Model Results}
    {%
    \begin{tabular}{lcccccccc}
        \hline
        Model & Accuracy & F1 Score & RMSE & RMSE \% & SP Mean & SP Max & EoO Mean & EoO Max \\
        \hline
        ft\_all & 0.612 & 0.510 & 1.428 & 52.40 & 0.034 & 0.147 & 0.071 & 0.174 \\
        \hline
    \end{tabular}
    }
\end{table}

The random forest model achieves a reasonable accuracy of 61.2\%, but its fairness metrics contain notable disparities. Specifically, it records the worst SP mean and SP max and the second worst EoO mean and max across all models.

Comparatively, the fine-tuned 70B LLM improved predictive performance, reaching 70.2\% accuracy. Nevertheless, disparity outcomes remain similar to those observed in the random forest model. However, with the addition of case note summaries, disparities reduced significantly while simultaneously improving accuracy, lowering EoO mean from 0.071 in the RF and base fine-tuned model to 0.021. This calls attention to the unique strength of LLMs to effectively incorporate contextual information in ways that mitigate disparities that otherwise exist in purely tabular settings.

\vspace{0.5em}
\textbf{Conclusions:} The random forest results in a comparable accuracy to the 70B LLM models, but also exhibits one of the largest SP and EoO disparities. However, when augmented with summaries, the 70B model can surpass RF on accuracy as well as reduce fairness gaps. This demonstrates a key advantage LLMs have compared to tabular models, as they can interpret unstructured context that cannot otherwise be as effectively incorporated.

\section{Feature Importances}

Another method used was identifying features using a Random Forest model trained on  tabular covariates. Through this, the feature with the highest importance score was included in the LLM prompt. Ultimately, the top feature was max placement before 2019. 

\begin{table}[ht]
    \centering
    \normalsize
    \renewcommand{\arraystretch}{1.1}
    \setlength{\tabcolsep}{2pt}
    \captionsetup{skip=3pt}
    \normalsize
    \renewcommand{\arraystretch}{1.1}
    \setlength{\tabcolsep}{2pt}
    \caption{Zero-Shot 17B Feature Importance Comparison}
    {%
    \begin{tabular}{lcccccccc}
        \hline
        Model & Accuracy & F1 Score & RMSE & RMSE \% & SP Mean & SP Max & EoO Mean & EoO Max \\
        \hline
        Base Prompt & 0.331 & 0.191 & 1.282 & 57.27 & 0.025 & 0.094 & 0.028 & 0.088 \\
        Base Prompt + Top Feature & 0.615 & 0.383 & 1.403 & 53.23 & 0.029 & 0.093 & 0.042 & 0.141 \\
        \hline
    \end{tabular}
    }
\end{table}

\begin{table}[ht]
    \centering
    \normalsize
    \renewcommand{\arraystretch}{1.1}
    \setlength{\tabcolsep}{2pt}
    \captionsetup{skip=3pt}
    \normalsize
     \renewcommand{\arraystretch}{1.1}
     \setlength{\tabcolsep}{2pt}
    \caption{Zero-Shot 17B + Summaries Feature Importance Comparison}
    {%
    \begin{tabular}{lcccccccc}
        \hline
        Model & Accuracy & F1 Score & RMSE & RMSE \% & SP Mean & SP Max & EoO Mean & EoO Max \\
        \hline
        Base Prompt & 0.470 & 0.331 & 1.160 & 61.32 & 0.026 & 0.074 & 0.044 & 0.125 \\
        Base Prompt + Top Feature & 0.597 & 0.392 & 1.287 & 57.10 & 0.030 & 0.108 & 0.032 & 0.155 \\
        \hline
    \end{tabular}
    }
\end{table}      

After adding the top feature to these models, we observe that accuracy improved significantly. This is expected, since explicitly including the most important feature provides the model with valuable information about which variable carries the strongest predictive signal and ultimately guiding its predictions more effectively.

However, in terms of fairness, the results are more variable. For the model without summaries, both EoO mean and EoO max increased, indicating greater disparities across groups and at least one subgroup becoming more disadvantaged. This suggests that although the max placement type before 2019 is not itself a protected attribute, it is correlated with characteristics such as race or gender due to systemic differences in housing access. Therefore, highlighting this feature therefore strengthens accuracy while simultaneously reinforcing historical biases.

In contrast, when case note summaries were combined with the top feature, EoO mean decreased relative to the summaries-only baseline, which shows that emphasizing this feature helped the model counteract some of the additional disparities introduced by summaries. At the same time, EoO max worsened, showing that at least one subgroup still experienced unequal treatment despite improvements in average disparity.

Overall, these results highlight an important interaction. Summaries are helpful for providing contextual information, but they also expand the input space and introduce noise, making the feature space more high-dimensional. In this setting, smaller models such as the 17B can be misguided by the additional variability and by providing an explicit feature importance prompt, it signals the model toward a highly predictive feature, improving average fairness in the presence of summaries even though it worsens fairness when used alone.

Given that model capacity can influence a model's predictions, we also analyzed how the feature importance cue interacts with the 70B model to determine whether the patterns observed hold at a larger scale.

\begin{table}[ht]
    \centering
    \captionsetup{skip=3pt}
    \normalsize
    \renewcommand{\arraystretch}{1.1}
    \setlength{\tabcolsep}{2pt}
    \caption{Fine-Tuned 70B Feature Importance Comparison}
    {%
    \begin{tabular}{lcccccccc}
        \hline
        Model & Accuracy & F1 Score & RMSE & RMSE \% & SP Mean & SP Max & EoO Mean & EoO Max \\
        \hline
        Base Prompt & 0.702 & 0.546 & 1.254 & 58.19 & 0.027 & 0.110 & 0.071 & 0.209 \\
        Base Prompt + Top Feature & 0.697 & 0.500 & 1.252 & 58.27 & 0.020 & 0.114 & 0.042 & 0.157 \\
        \hline
    \end{tabular}
    }
\end{table}

\begin{table}[ht]
    \centering
    \captionsetup{skip=3pt}
    \normalsize
    \renewcommand{\arraystretch}{1.1}
    \setlength{\tabcolsep}{2pt}
    \caption{Fine-Tuned 70B + Summaries Feature Importance Comparison}
    {%
    \begin{tabular}{lcccccccc}
        \hline
        Model & Accuracy & F1 Score & RMSE & RMSE \% & SP Mean & SP Max & EoO Mean & EoO Max \\
        \hline
        Base Prompt & 0.686 & 0.504 & 1.252 & 58.26 & 0.025 & 0.081 & 0.021 & 0.097 \\
        Base Prompt + Top Feature & 0.686 & 0.519 & 1.233 & 58.89 & 0.032 & 0.140 & 0.060 & 0.470 \\
        \hline
    \end{tabular}
    }
\end{table}

Compared to the 17B model, the 70B model without summaries reveals the opposite pattern where performance metrics decreased slightly, but fairness improved with reductions in both EoO mean and max. This indicates that post fine-tuning, the larger model already learned to effectively weight the most predictive features, making the explicit cue less beneficial for accuracy.

When summaries were added, however, the 70B model’s behavior differed from the 17B model's. In the baseline with summaries, subgroup disparities such as EoO mean and max are much smaller than in the non-summary setting, suggesting that the additional contextual information helps the model smooth its predictions across groups. However, once feature importance was added to the summaries, fairness deteriorated considerably with the EoO max rising to almost four times its original value. This likely reflects the fact that the fine-tuned 70B has established an internal balance for weighting features. By disrupting that balance through prompting it to focus on a single historically predictive feature, it negatively affected fairness. Thus, while summaries in a larger fine-tuned model can mitigate average disparities, overemphasizing one feature can override those benefits and reintroduce bias.

\begin{table}[H]
    \centering
    \captionsetup{skip=3pt} 
    \caption{Top Feature vs Layering Additional Features Metrics}
    \renewcommand{\arraystretch}{1.1} 
    \setlength{\tabcolsep}{8pt} 
    \begin{tabular}{lccc}
        \hline
        \textbf{Metric} & \textbf{Top Feature} & \textbf{Min (V2--V11)} & \textbf{Max (V2--V11)} \\
        \hline
        SP Mean & 0.03 & 0.03 & 0.036 \\
        SP Max & 0.108 & 0.076 & 0.112 \\
        EoO Mean & 0.032 & 0.036 & 0.051 \\
        EoO Max & 0.155 & 0.151 & 0.170 \\
        \hline
    \end{tabular}
\end{table}

Additionally, we evaluated how incrementally adding more features by importance to the Zero-Shot 17B + Summaries model (V2–V11) impacted fairness by comparing disparity metrics to the case where only the top predictor was used, considering up to 10 additional features at once. As shown in Table 8, the fairness metrics for the top feature alone are comparable to, and in some cases slightly better than, those obtained when additional features are layered in.

 This pattern indicates that simply adding more features does not guarantee improved fairness and, in some cases, can even amplify subgroup disparities.

\vspace{0.5em}
\textbf{Conclusions:} Overall, prompting with the top tabular feature raises zero-shot 17B accuracy, but may worsen fairness due to correlation between that feature and historically unequal access. Conversely, with summaries the average fairness can improve while worst-case gaps may continue to widen. In the fine-tuned 70B, explicitly emphasizing the top feature leads to little accuracy gain and may disrupt the model’s learned balance. Furthermore, results show that adding more features (V2–V11) does not reliably improve fairness as opposed to the single top feature.

\section{True Positive Rate Bar Plots}

\begin{figure}[H]
    \centering
    \includegraphics[width=\textwidth]{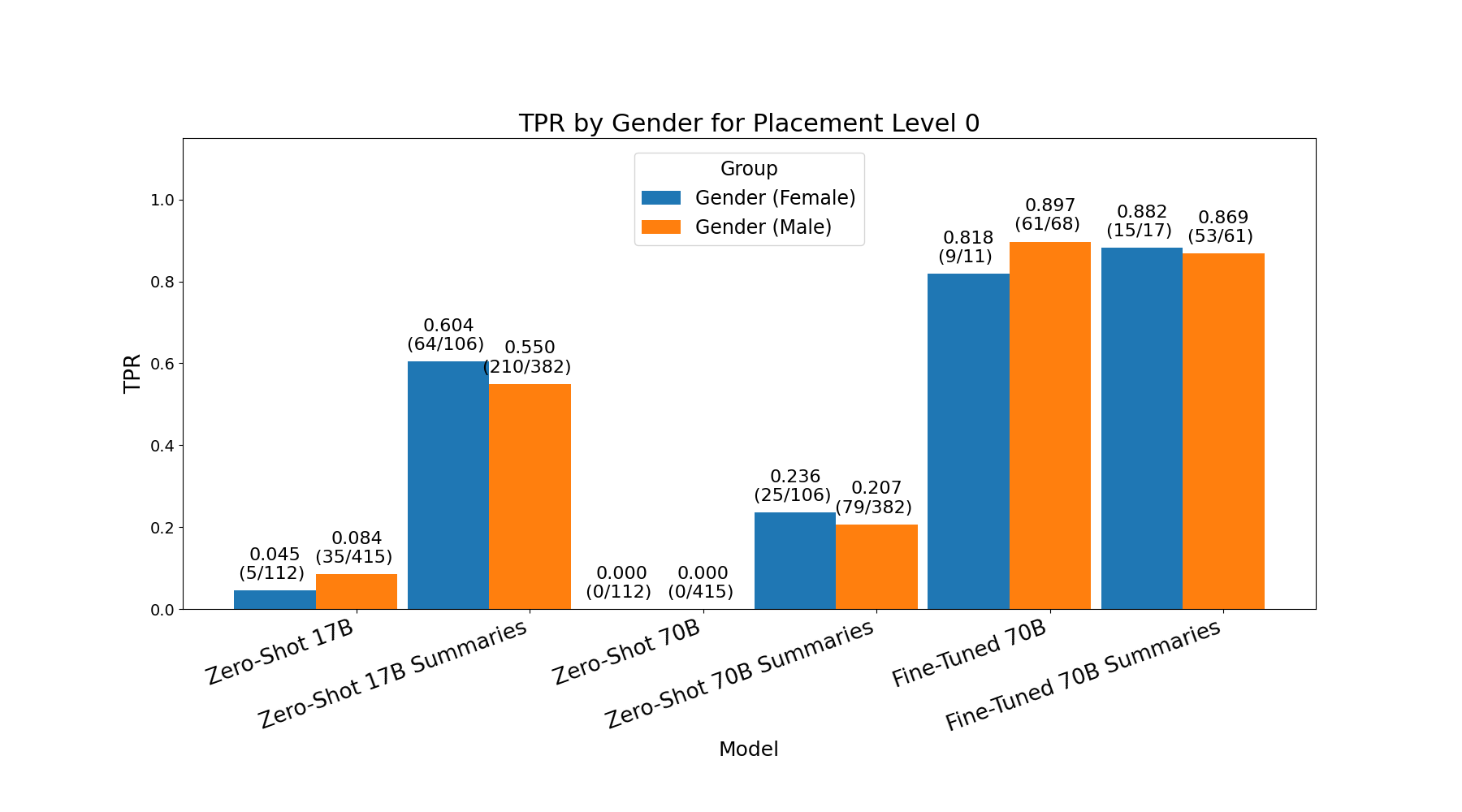} 
    \caption{True Positive Rate (TPR) Bar Plot Level 0 (Gender)}
\end{figure}

\begin{figure}[H]
    \centering
    \includegraphics[width=\textwidth]{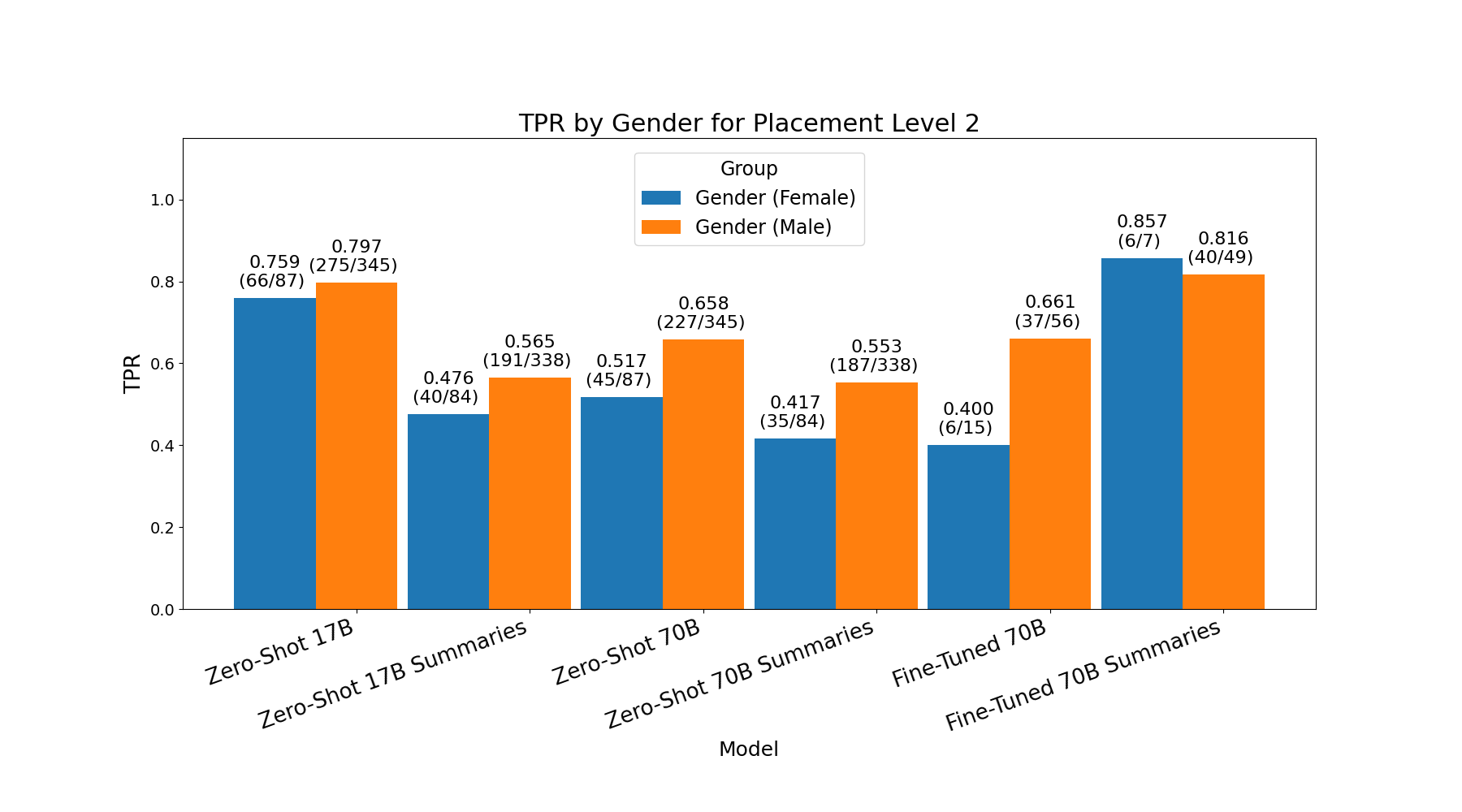} 
    \caption{True Positive Rate (TPR) Bar Plot Level 2 (Gender)}
\end{figure}

\begin{figure}[H]
    \centering
    \includegraphics[width=\textwidth]{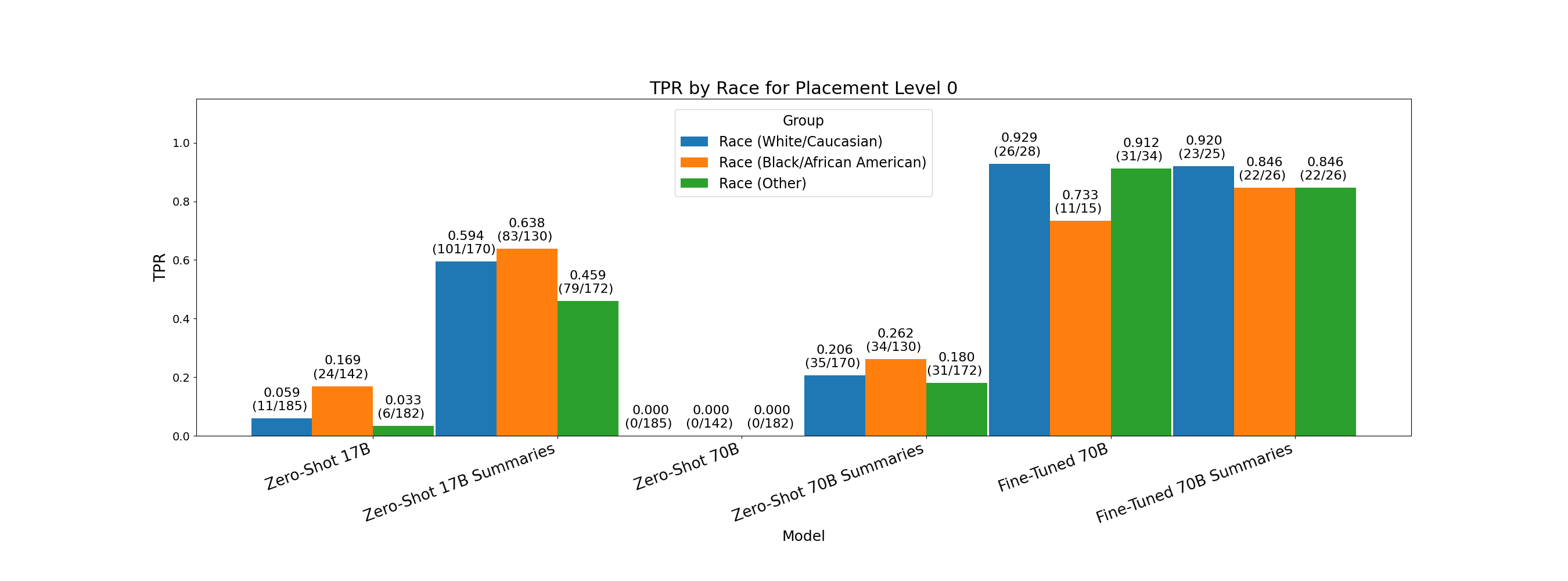} 
    \caption{True Positive Rate (TPR) Bar Plot Level 0 (Race)}
\end{figure}

\begin{figure}[H]
    \centering
    \includegraphics[width=\textwidth]{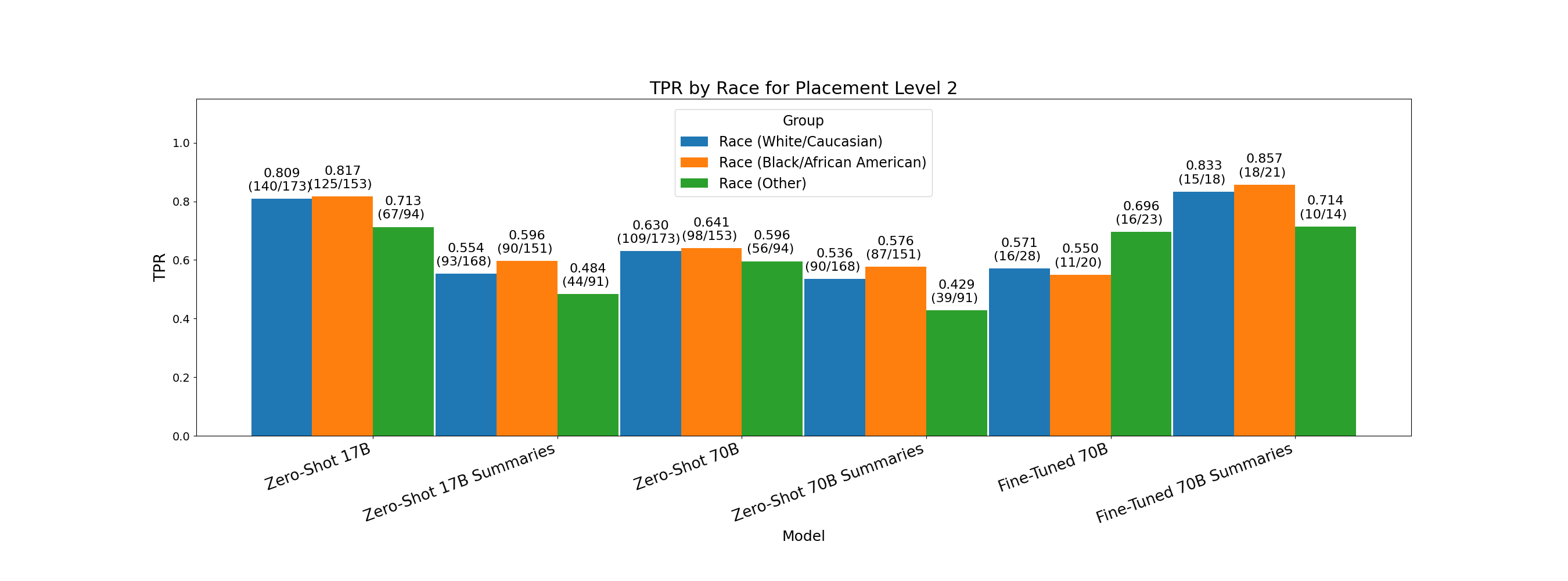} 
    \caption{True Positive Rate (TPR) Bar Plot Level 2 (Race)}
\end{figure}

We also plotted the TPR for certain protected subgroups. From the above plots, various patterns can be observed. First, female TPR is consistently lower than male TPR across most models, with the disparity particularly pronounced in the fine-tuned model. This is likely attributable to the dataset’s skewed gender distribution, which at times exceeded a 4:1 ratio of men to women, so the model is exposed to more male examples and learns to predict outcomes for the majority group more effectively.

Second, the zero-shot models tend to favor predicting class 2. This behavior could indicate that the task is relatively complex for  the pre-trained base models, that the prediction pattern aligns more closely with their pre-training data, or that the models are effectively making a “safer” middle-class prediction to reduce overall error. The noticeable improvement in TPR for placement level 0 after fine-tuning supports the idea that the pre-trained base models may struggle with this task until it is specialized to our application.

Finally, we observe that adding summaries appears to help mitigate some gender disparities, and compared to gender, racial disparities are less pronounced overall. This could be due to the more uniform distribution of samples across racial groups, compared to the skewed gender distribution.

\vspace{0.5em}
\textbf{Conclusions:} Female TPRs are generally lower than male TPRs, with the disparity being the most evident in fine-tuned models, which is likely due to the underlying data imbalance. However, summaries tend to narrow gender gaps. Additionally, zero-shot models often gravitate towards mid-level predictions (class 2 in our case). Notably, racial disparities are less pronounced than gender in our setting.

\section{Overall Conclusions and Recommendations}

Our first conclusion is that despite widely documented biases in word embeddings and LLM predictions in sensitive settings, the casenotes dataset in this study typically comprises of short casenotes that do not appear to introduce additional algorithmic biases from text beyond potential biases in tabular classification tasks. In explorations not documented here, when asked for reasoning or justification for tabular predictions, the LLMs did not mention race, gender or sensitive attributes. The casenotes themselves are heavily redacted and brief. Our overall findings regarding textual biases may not extend to other complex settings where case-notes are written in human services, especially if notes are more narrative or longer than in our dataset.

Imbalanced datasets can introduce disparities in improvements from fine-tuning. Despite having the highest accuracy due to the optimization process, the fine-tuned 70B model displayed some of the worst fairness outcomes. Especially for minority groups such as women, who were represented by less than a quarter the number of men, the disparate impacts of fine-tuning were particularly evident.

Another key finding is the impacts of case note summaries. Given the rich background summaries provide, summaries can aid in improving accuracy for zero-shot models. However, in the same note, the additional information can also introduce noise that distracts larger models. Despite being less effective performance-wise in fine-tuned models, incorporating summaries utilizes the model’s deeper understanding of nuanced language and shows potential to improve fairness for disadvantaged groups.

Finally, feature importance guidance can be beneficial for zero-shot models. Since the model does not have any specialized training, embedding the most important feature into the prompt leads to additional emphasis on that feature during prediction. However, in a fine-tuned model, this approach is not as effective. Through training, the model has learned statistically appropriate weights and emphasizing one feature can lead to negatively affecting that balance. Especially if the feature reflects historical bias, it may be prone to amplifying disparities. 

Overall, fairness metrics must continue to be carefully monitored to ensure that disparities are not overlooked or exacerbated as predictive performance improves.

\paragraph{Acknowledgments}
We thank extensive discussions with our nonprofit partners and Lauri Goldkind. We thank the USC Jumpstart program and acknowledge the DSO Summer Scholars program for support of Xiao Qi Lee's undergraduate research. We thank Together AI for providing exploratory cloud compute credits. 

\bibliographystyle{plainnat}
\bibliography{refs}

\appendix
\setcounter{figure}{0}
\renewcommand{\thefigure}{\arabic{figure}}
\section*{Appendix}\label{app:confusion_matrices}

\begin{figure}[H]
  \centering
  \includegraphics[width=0.8\textwidth]{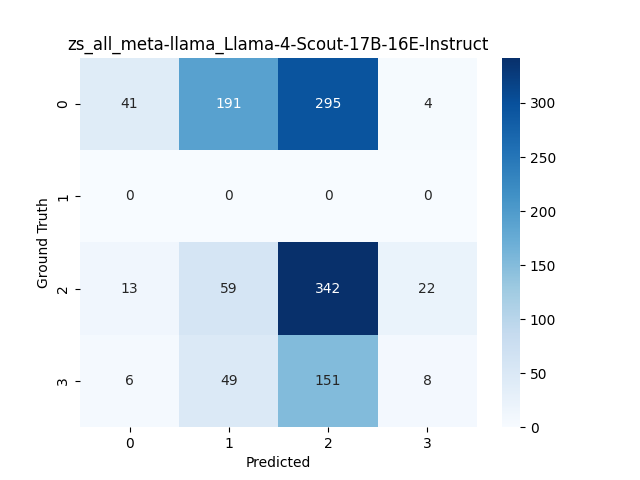}
  \caption{Confusion Matrix for Zero-Shot 17B}
  \label{fig:cm_zs17b}
\end{figure}

\begin{figure}[H]
  \centering
  \includegraphics[width=0.8\textwidth]{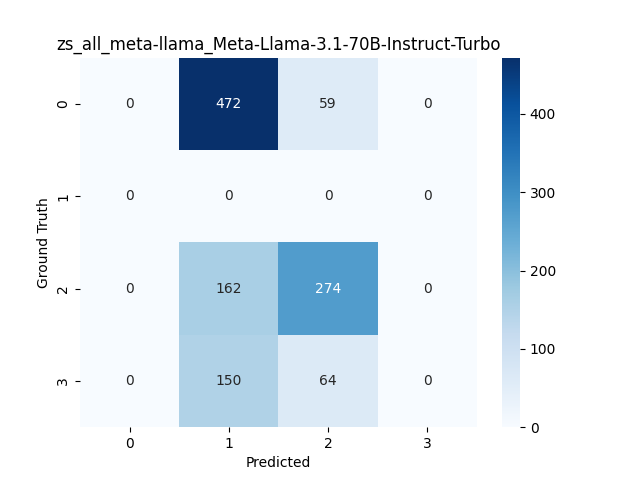}
  \caption{Confusion Matrix for Zero-Shot 70B}
  \label{fig:cm_zs70b}
\end{figure}

\begin{figure}[H]
  \centering
  \includegraphics[width=0.8\textwidth]{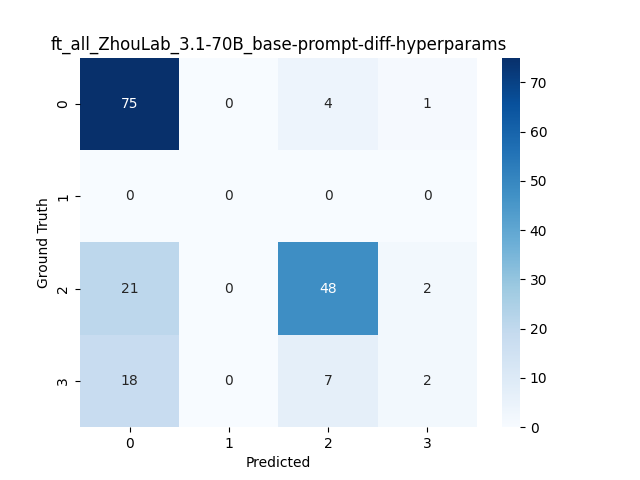}
  \caption{Confusion Matrix for Fine-Tuned 70B}
  \label{fig:cm_ft70b}
\end{figure}

\begin{figure}[H]
  \centering
  \includegraphics[width=0.8\textwidth]{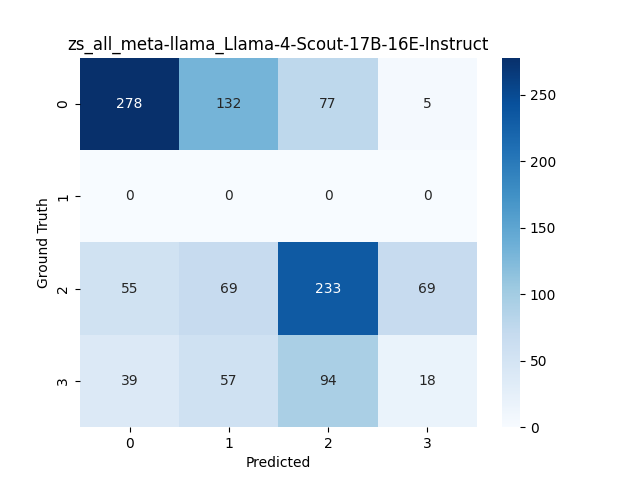}
  \caption{Confusion Matrix for Zero-Shot 17B + Summaries}
  \label{fig:cm_zs17b_summ}
\end{figure}

\begin{figure}[H]
  \centering
  \includegraphics[width=0.8\textwidth]{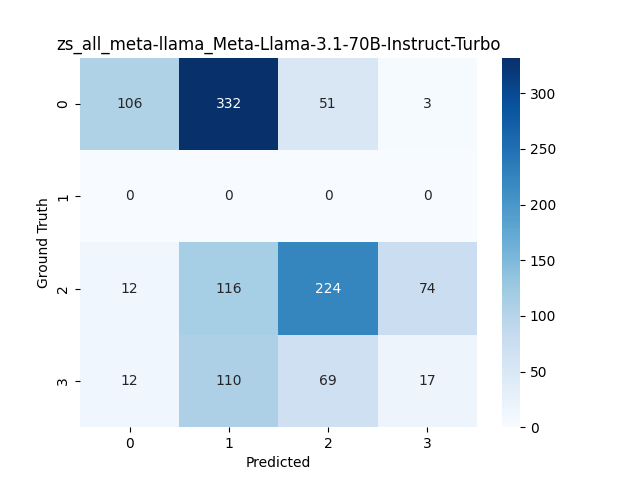}
  \caption{Confusion Matrix for Zero-Shot 70B + Summaries}
  \label{fig:cm_zs70b_summ}
\end{figure}

\begin{figure}[H]
  \centering
  \includegraphics[width=0.8\textwidth]{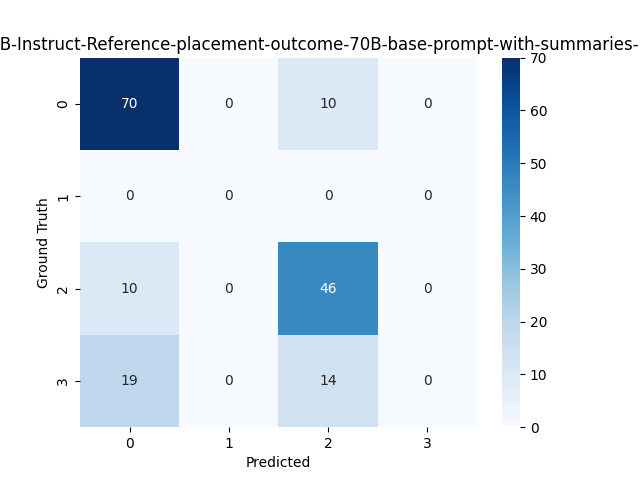}
  \caption{Confusion Matrix for Fine-Tuned 70B + Summaries}
  \label{fig:cm_ft70b_summ}
\end{figure}

\end{document}